\pdfoutput=1
\documentclass[pdftex,12pt]{JHEP3}
\usepackage{amsfonts,amsbsy,latexsym,shadow,fancybox,amssymb,amscd,amstext,amsmath}
\usepackage[pdftex]{epsfig}
\pdfoutput=1

\renewcommand{\AmS}{{\protect\the\textfont2A\kern-.1667em\lower.5ex\hbox{M}\kern-.125emS}}
 \def\be{\begin{equation}}
\def\ee{\end{equation}}
\def\bea{\begin{eqnarray}}
\def\eea{\end{eqnarray}}
\def\pd{\partial}
\def\a{\alpha}
\def\b{\beta}
\def\g{\gamma}
\def\d{\delta}
\def\m{\mu}
\def\n{\nu}

\def\l{\lambda}

\def\r{\rho}
\def\s{\sigma}

\def\bg{\bar{g}}
\def\bn{\bar{\nabla}}

\def\bg{\bar{g}}

\def\bn{\bar{\nabla}}
\def\bi{\begin{itemize}}
\def\ei{\end{itemize}}
\def\bn{\bar{\nabla}}
\def\bR{\bar{R}}

\date{ 2011} \preprint{IFT-UAM/CSIC-12-89;FTUAM-12-102}
\title{Unimodular gravity with external sources.} \author{Enrique \'Alvarez and Mario Herrero-Valea \\  Instituto de F\'{\i}sica Te\'orica
UAM/CSIC and Departamento de F\'{\i}sica Te\'orica \\ Universidad
Aut\'onoma de Madrid, E-28049--Madrid, Spain \\ E-mail: \email{enrique.alvarez@uam.es\\mario.herrero@estudiante.uam.es }}
\abstract{
The only allowed source of the gravitational field in the unimodular theory, invariant under area-preserving (transverse) diffeomorphisms as well as Weyl transformations, is just the traceless piece of the energy-momentum tensor.
This fact notwithstanding, the  free energy produced by  arbitrary sources (not only static ones) is identical to the one predicted by general relativity. This encompasses all weak field tests of gravitation.

}
\begin{document}

{\vskip 1cm}
\newpage
\vskip 1cm

\vskip 1cm


\newpage
\tableofcontents


\setcounter{page}{1}
\setcounter{footnote}{1}
\tableofcontents

\section{Introduction}
The purpose of the present paper is to study the basic unimodular gravitational action (U from now on), which in the Einstein frame corresponds to a unimodular metric, with determinant 
\be
g_E\equiv \text{det}~g^E_{\m\n}=1
\ee
(in Lorentzian signature we always write $g$ to mean the positive number $|g|$). The coupling to matter fields has been worked out in the second reference of \cite{Alvarez} using Weyl transverse symmetry (WTdiff) as a guiding principle \footnote{
This theory was first discovered in a flat space setting in \cite{AlvarezBGV} in the context of a general analysis of TDiff invariant theories. Those are invariant under transverse (area preserving) diffeomorphisms only. The first such theory is due to Einstein himself \cite{Einstein}. Other relevant references are \cite{Bij}\cite{Unruh}\cite{Wilczek}\cite{Ellis}\cite{Hooft}.
}. Its main characteristic is that the potential energy does not couple directly to the gravitational field. In general relativity (GR) this coupling is through the square root of the determinant of the metric, which in our case is set to unity in Einstein's frame. 

\noindent This seems promising as a solution to the direct gravitational constant problem, namely why vacuum energy seems to weigh much less that other forms of energy as if the equivalence principle did not hold for it. This is logically independent of the question as to why the universe is expanding in an accelerated way.
Nevertheless, as we shall see in detail in the following, the interaction of arbitrary external sources is exactly the same as in GR.
\par
\noindent The simplest action in this setting, using a scalar field as an example, reads
\bea
&&S\equiv S_g+S_m=\int d^n x \left( -M^{n-2} R_E+{1\over 2} g_E^{\a\b}\pd_\a\phi\pd_\b\phi-V(\phi)\right)=\nonumber\\
&&- M^{n-2}\int d^n x \; g^{1\over n}\left[\left(R+{n-1\over n}{\nabla^2 g\over g}-{(n-1)(5 n-2)\over 4 n^2}{(\nabla g)^2\over g^2}\right)\right]+\nonumber\\
&&+\int d^n x \; \left[g^{1\over n}\left( {1\over 2}g^{\a\b}\pd_\a\phi\pd_\b\phi\right)-V(\phi)\right]\\
\eea
In the second formula we have written the action in an arbitrary frame, using
\be
g^E_{\m\n}=g^{-{ 1\over n}}~g_{\m\n}
\ee
 It is to be stressed that this is {\em not} a field redefinition, because it is not invertible. 
 Actually,
 \be
 \d g^E_{\m\n}\equiv M_{\m\n}^{\a\b} \d g_{\a\b}=g^{-{1\over n}}\left({1\over 2}\left(\d^\a_\m\d^\b_\n+\d^\a_\n\d^\b_\m\right)-{1\over n}g^{\a\b} g_{\m\n}\right)\d g_{\a\b}
 \ee
 It is not possible to recover the generic metric out of Einstein's unimodular metric (the converse is of course trivial). This action in an arbitrary frame is then invariant under local Weyl transformations
 \be
 \tilde{g}_{\m\n}=e^{2\s(x)} g_{\m\n}
 \ee
 as well as under area preserving diffeomorphisms (id est, those with unit jacobian determinant). This is precisely the symmetry that was dubbed WTDiff in previous papers.
 \par
 The gravitational part of the equations of motion (EM) can be written in the manifestly traceless form (which is much more convenient than the one in \cite{Alvarez}). Instead of deriving them from the variational principle (which we also did) it is much better to use the fact that the Euler-Lagrange equations behave as forms in configuration space \cite{Arnold}. The Jacobian factor just instructs us to take the trace after performing the point transformation on Einstein's equations. This yields in general
 \be
 R_{\a\b}-{1\over n} R g_{\a\b}+(n-2)\left(\nabla_\a\nabla_\b\s-\nabla_\a\s\nabla_\b\s\right)+{n-2\over n}\left(\nabla^2\s-\left(\nabla\s\right)^2\right)g_{\a\b}=0
 \ee
Using now the fact that in our case
\be
e^{2\s}=g^{-{1\over n}}
\ee
and including also the contribution of the matter field the final result is gotten 
 \bea
 &&R_{\m\n}-{1\over n}~R~g_{\m\n}\equiv ~M^{2-n} \left(J^g_{\m\n}+J^m_{\m\n}\right)=\nonumber\\
 &&{(n-2)(2n-1)\over 4 n^2}\left({\nabla_\m g\nabla_\n g\over g^2}-{1\over n}{(\nabla g)^2\over g^2} g_{\m\n}\right)-{n-2\over 2n}\left({\nabla_\m\nabla_\n g \over g}-{1\over n}{\nabla^2 g\over g} g_{\m\n}\right)+\nonumber\\
 &&+{M^{2-n}\over 2}\left(\pd_\m \phi\pd_\n\phi-{1\over n}~g^{\a\b}\pd_\a\phi\pd_\b\phi~ g_{\m\n}\right)
\eea 
It can be checked that the EM are Weyl invariant, as they should be by construction.\vspace{0.4cm}
We have included in the source, besides the matter part, $J^m_{\m\n}$, a gravitational piece $J^g_{\m\n}$ coming from the space-time dependence of the metric determinant.
This means that we are defining the sources in the matter part
\be
g^{1\over n}~J_m^{\m\n}\equiv {\d S_m \over \d g_{\m\n}}
\ee
as well as in  the gravitational piece
\be
g^{1\over n}\left(R_{\m\n}-{1\over n} ~R~g_{\m\n}+M^{2-n}~J^g_{\m\n}\right)\equiv {\d S_g \over \d g_{\m\n}}
\ee
A useful identity can be also obtained by taking the contracted covariant derivative and using the contracted Bianchi identity corresponding to the Weyl-transformed Einstein tensor
\be
\tilde{\nabla}_\m \tilde{S}^{\m\n}\equiv \tilde{\nabla}_\m \left(\tilde{R}^{\m\n}-{1\over 2}\tilde{R}_\m \tilde{g}^{\m\n}\right)=0
\ee
which yields
\be
\tilde{\nabla}_\m \left(\tilde{R}^{\m\n}-{1\over n}\tilde{R}\tilde{g}^{\m\n}\right)={n-2\over 2n}\tilde{\nabla}^\n \tilde{R}
\ee
\par

\par In this work we shall stick to the standard definition of energy-momentum tensor in general relativity (GR) which for a scalar field with minimal coupling  means
\be
T_{\m\n}\equiv \pd_\m \phi\pd_\n\phi-\left({1\over 2}g^{\a\b}\pd_\a\phi\pd_\b\phi-V(\phi)\right)g_{\m\n}
\ee
Its trace is
\be
T\equiv g^{\a\b} T_{\a\b}=n V-{n-2\over 2}\left(\nabla\phi\right)^2
\ee
so that the piece of the source that depends on the scalar field is precisely
\be
J^m_{\m\n}=T_{\m\n}-{1\over n}~T~g_{\m\n}=\pd_\m\phi\pd_\n\phi-{1\over n}~g^{\a\b}\pd_\a\phi\pd_\b\phi~g_{\m\n}
\ee
and does not include the potential energy. Thus it seems plain that
\be
\nabla_\m \left(T^{\m\n}-{1\over n}~T~g^{\m\n}\right)=Ð{1\over n}\nabla^\n T
\ee

We shall see in a moment that this is so in general precisely owing to the Ward identities of the area-preserving diffeomorphisms plus Weyl symmetry.
To be specific, Ward identities guarantee that there is a function $\Theta$ such that
\be
\nabla_\m \left(g^{{1\over n}-{1\over 2}} {\d S\over \d g _{\m\n}}\right)=\nabla^\n \Theta
\ee

\par

We have to be careful however. The variation of the action under a variation of the scalar field reads
\bea
\d S_m&&=\int d^n x g^{1\over n} g^{\m\n}\pd_\n \d\phi-V^\prime \d\phi=-\int d^n x~\left(\pd_\n\left(g^{1\over n}~g^{\m\n}\pd_\m\phi\right)+V^\prime\right)\d\phi=\nonumber\\
&&=\int d^n x g^{1\over n}\left(\nabla^2\phi-{n-2\over 2n}\nabla\phi.{\nabla g\over g}\right)+V^\prime (\phi)
\eea
It follows that the EM of the scalar field have changed; now they read
\be
\nabla^2 \phi +g^{-{1\over n}} V^\prime = {n-2\over 2n}{\nabla \phi.\nabla g\over g}
\ee
It is then not difficult to check that
\bea
&&\nabla_\m\left(g^{{1\over n}-{1\over 2}}\left(\nabla^\m\phi\nabla^\n\phi-{1\over n}(\nabla\phi)^2 g^{\m\n}\right)\right)=\nonumber\\
&&g^{2-n\over 2n}\nabla^\n\left(V+{n-2\over 2n}\left(\nabla\phi\right)^2 \right)=g^{2-n\over 2n}{1\over n}\nabla^\n T
\eea
which modulo EM implies the desired result. Indeed, for any arbitary constant value of $\l$,
\be
\nabla_\m \left(g^\l {\d S\over \d g_{\m\n}}\right)=\left(\nabla_\m g^\l\right){\d S\over \d g_{\m\n}}+g^\l\nabla_\m {\d S\over \d g_{\m\n}}
\ee
the extra term is multiplied by the EM. 
\vspace{1cm}

In this paper we would like to restrict our attention to the quadratic action corrresponding to fluctuations around a given flat background coupled to arbitrary external sources. A detailed  analysis of the equations of motion will be performed in order to compare them to those of GR. The main conclusion of our work is that, although the only allowed source of the gravitational field in the unimodular theory is just the traceless piece of the energy-momentum tensor
\be
J_{\m\n}\equiv T_{\m\n}-{1\over n} T g_{\m\n}
\ee
the {\em full free energy} produced by {\em arbitrary} sources (not only static ones) is {\em identical} to the one predicted by general relativity. This encompasses all weak field tests of gravitation.
\par
Then we analyze the corresponding equations of motion corresponding to perturbations around an arbitrary background. We are able to show that the unimodular ones are a subset of the ones corresponding to general relativity, with the source restricted to its traceless piece and with vanishing cosmological constant.
This refines a previous analysis by one of us \cite{Alvarez}. There it was argued that Bianchi identities on the unimodular theory just imply a first integral of the equations of motion which, once used, made the unimodular theory fully equivalent to GR with an arbitrary cosmological constant. Here we claim a stronger result, namely that at the level of external sources, this cosmological constant must vanish.
\par

\section{Ward identities.}
Let us first review the steps necessary to derive Bianchi identities from the field theory viewpoint \cite{Alvarez}. We perform a transformation of symmetry in the action
\be
\d S=\int d^n x \sqrt{g}~\d g_{\a\b}~{1\over \sqrt{g}}{\d S\over \d g_{\a\b}}
\ee
This should vanish identically whenever the variation of the metric is generated by a diffeomorphism
\be
\d g_{\a\b}=\nabla_\a \xi_\b+\nabla_\b \xi_\a
\ee
that is
\be
0=\int d^n x \sqrt{g}~\left(\nabla_\a \xi_\b+\nabla_\b \xi_\a\right)~{1\over \sqrt{|g||}}{\d S\over \d g_{\a\b}}=-2\int d^n x \sqrt{|g|}\xi_\a \nabla_\a {1\over \sqrt{|g|}}{\d S\over \d g_{\a\b}}
\ee
The fact that the vector $\xi_\a$ is arbitrary conveys the fact that 
\be
\nabla_\a {\d S\over \d g_{\a\b}}\equiv 0
\ee
{\em any} covariant lagrangian leads to such an identity, which in the case of Einstein-Hilbert is the contracted Bianchi identity

\be
\nabla_\a\left(R^{\a\b}-{1\over 2}~R~g^{\a\b}\right)\equiv 0
\ee

\par
In order to restrict to area preserving diffeomorphisms, let us now impose transversality on the generator
\be
\nabla_\a\xi^\a=0
\ee

It is useful to define a one form $\xi_1\equiv \xi_\m dx^\m$ as well as a two form $\Omega_2\equiv{1\over 2}\Omega_{\m\n}~dx^\m\wedge dx^\n$. The transversality condition now reads
\be
\xi_1=-2\d\Omega_2
\ee
where the codifferential is the adjoint operator of the exterior derivative. Acting on two-forms
\be
\d\equiv *^{-1}~d~*
\ee
so that its components obey 
\be
\left(\d\Omega_2\right)_\r=-{1\over 2}\nabla^\n \left(\Omega_2\right)_{\n\r}
\ee
The number of independent components of a two form is clearly ${n\choose 2}$, but the codifferential is nilpotent $\d^2=0$, so that we have to withdraw the three forms
\be
\Omega_2=\d \Omega_3
\ee
and from them we have got to withdraw the four forms, etc. The final counting of undependent gauge parameters is:
\be
{n\choose 2}-\left({n\choose 3}-\left({n\choose 4}-\left(\ldots \right)\right)\right)=n-1
\ee

where the relationship $\sum_j \left(-1\right)^j~{n\choose j}=0$ has been used.

Taking into account that for antisymmetric tensors $\Omega^{\a\b}=-\Omega^{\b\a}\Rightarrow \nabla_\a\nabla_\b \Omega^{\a\b}\equiv 0$
\bea
&&0=\int d^n x ~\left(\nabla_\a \xi_\b+\nabla_\b \xi_\a\right)~{\d S\over \d g_{\a\b}}=\int d^n x ~\left(\nabla_\a \nabla^\r \Omega_{\b\r}+\nabla_\b \nabla^\r\Omega_{\a\r}\right)~~{\d S\over \d g_{\a\b}}=\nonumber\\
&&-2 \int d^n x \sqrt{|g|}~\Omega^{\r\b}~ \nabla^\r\nabla_\a ~{1\over \sqrt{|g|}}~{\d S\over \d g_{\a\b}}
\eea
Assuming that $\Omega^{\r\b}$ is arbitrary (and as we have just seen, they are either arbitrary or else vanishing) we deduce that
\be
\nabla^\r\nabla_\a ~{1\over \sqrt{g}}~{\d S\over \d g_{\a\b}}=\nabla^\b\nabla_\a ~{1\over \sqrt{|g|}}~{\d S\over \d g_{\a\r}}
\ee
that is for the vector
\be
\Theta^\b\equiv \nabla_\a ~{1\over \sqrt{g}}~{\d S\over \d g_{\a\b}}
\ee
the condition
\be
\nabla^\r\Theta^\b=\nabla^\b\Theta^\r
\ee
which can be integrated as
\be
\Theta_\r=\nabla_\r \Phi+\g_\r
\ee
where $\g\equiv \g_\r dx^\r$ is an harmonic form. The number of independent such things depends on the topology of the manifold, and is referred to as the first betti number, $b_1(M)$, the dimension of the first cohomology group, $H^1(M)$.
\par
In case there are not harmonic forms in the space-time manifold,(which happens, in particular, if it is diffeomorphic to $\mathbb{R}^n$), this shows that the Bianchi identity is modified in the sense that
\be
\nabla^\r\Theta^\b=\nabla^\b\Theta^\r=\nabla^\b\nabla^\r \Phi
\ee

{\em id est}, still hold when integrated over the whole of space-time with the Diff invariant measure.
The Weyl invariance of the action means that
\be
0=\int d^n x~ w(x)~g_{\a\b}~{\d S\over \d g_{\a\b}}
\ee
which conveys the fact that barring topological subtleties the trace of the EM must be a total derivative
\be
g_{\a\b}~{\d S\over \d g_{\a\b}}=\pd_\r\Sigma^\r
\ee
In order to get Ward identities out of symmetries of the action, we also need the parameters to be independent.
The standard method starts with a change of variables in the path integral expressing the expectation value of a certain monomial of fields, $X\left[g_{\m\n},\psi_i\right]$. where $\psi_i$ is a generic representation of matter fields.
\be
Z \left\langle 0_+| X\left[g,\psi_i\right]|0_- \right\rangle\equiv e^{i W}\equiv \int {\cal D}g_{\m\n}~{\cal D}\psi~X[g_{\m\n},\psi_i]~e^{i S_{grav}[g]+i S_{matt}\left[g,\psi_i\right]}
\ee
Namely $g_{\m\n}\rightarrow g_{\m\n}+\nabla_\m \xi_\n +\nabla_\n\xi_\m$, this leads easily to the Ward identity
\begin{align}
i \left\langle 0_+\left|{\d X\left[g,\psi_i\right] \over \d \Omega^{\m\n}(x)}\right|0_- \right \rangle=\nabla_\m \nabla^\a \left\langle  0_+\left|~X\left[g,\psi_i\right]~{ \d \tilde{S}\over \d g^{\a\n}(x)}\right|0_-\right \rangle-(\m \leftrightarrow \n)
\end{align}

In the particular case $X=1$, it states that the expectation value of the classical identity should vanish. There may be quantum corrections to the naive identities, either in the form of anomalies \cite{Jack} or even limit cycles \cite{Fortin}.


\section{Free energy with external sources in a flat background.}

The purpose of this section is to study the free energy of the unimodular theory neglecting self-interaction but in the presence of external sources. The result to be demonstrated is that it is fully equivalent with the general relativistic one.
\par
Let us start by consider perturbations around a flat background
\be
g_{\m\n}\equiv \eta_{\m\n}+\kappa h_{\m\n}
\ee
up to quadratic order.
In momentum space the kinetic term of the unimodular theory \cite{AlvarezBGV} reads\footnote{
We shall compare all the time with the GR template which at this order corresponds to the Fierz-Pauli (FP) spin two theory. Its kinetic energy piece reads
\bea
&&8 K_{FP}^{\m\n\r\s}=k^2\left(\eta^{\m\r}\eta^{\n\s}+\eta^{\m\s}\eta^{\n\r}-2\eta^{\m\n}\eta^{\r\s}\right)\nonumber\\
&&-\left(k^\m k^\r \eta^{\n\s}+k^\n k^\s \eta^{\m\r}+k^\m k^\s \eta^{\n\r}+k^\n k^\r \eta^{\m\s}-2 k^\m k^\n \eta^{\r\s}-2 k^\r k^\s \eta^{\m\n}\right)\nonumber
\eea
}
\bea
&&K^{U}_{\m\n\r\s}={1\over 8}k^2\left(\eta_{\m\r}\eta_{\n\s}+\eta_{\m\s}\eta_{\n\r}\right)-{1\over 8}\left(k_\n k_\s \eta_{\m\r}+\eta_{\m\s}k_\n k_\r+k_\n k_\r\eta_{\m\s}+k_\n k_\s\eta_{\m\r}\right)+\nonumber\\
&&{1\over 2 n}\left(\eta_{\m\n}k_\r k_\s+\eta_{\r\s}k_\m k_\n\right)-{n+2\over 4 n^2}k^2\eta_{\m\n}\eta_{\r\s}
\eea

This can be easily expressed in terms of the Barnes-Rivers projectors\footnote{The well-known ADM \cite{Arnowitt} formalism could be used just as well.} (reviewed in the appendix)
\bea
&&K^{U}={k^2\over 8}\left(2 P_2+ 2 P_0^s+ 2 P_1+ 2 P_0^w\right)-{k^2\over 8}\left(2 P_1 + 4 P_0^w\right)+\nonumber\\
&&{k^2\over 2 n}\left(\sqrt{n-1}P_0^\times+ 2 P_0^w\right)-{n+2\over 4 n^2}k^2 \left(\left(n-1\right)P_0^s+\sqrt{n-1}P_0^\times + P_0^w\right)=\nonumber\\
&&k^2\bigg\{{1\over 4}P_2-{n-2\over 4 n^2} P_0^s-{ n^2-3 n +2\over 4 n^2}P_0^w+{n-2\over 4 n^2}\sqrt{n-1}P_0^\times\bigg\}
\eea

It is plain that
\bea
&&K^{WT}_{\m\n\r\s}\eta^{\r\s}=0\nonumber\\
&& \xi.k=0~\Rightarrow~K_{\m\n\r\s}\xi^\r k^\s=0
\eea

Following \cite{AlvarezBGV}, we are going to gauge fix the unimodular theory by adding a gauge fixing lagrangian
\be
L_{gf}=h_{\m\n}~K_{gf}^{\m\n\r\s}~h_{\r\s}
\ee
where
\be
K_{gf}={k^6\over 4 \Lambda^4}P_1
\ee
This corresponds in position space to a gauge fixing
\be
L_{gf}={1\over 2 \Lambda^4} F_\a^2\equiv {1\over 2 \Lambda^4}\left(\pd_\a\pd^\m\pd^\n h_{\m\n}-\Box \pd^\m h_{\a\m}\right)^2
\ee
where $\Lambda$ is an arbitrary mass scale. Let us remark that this gauge choice is admissible, because it can be reached uniquely through area-preserving diffeomorphisms
\bea
&&\pd_\a F^\a=0\nonumber\\
&&\d F_\a=-\Box^2 \xi_\a
\eea

The ghost system associated to it gets complicated owing to the fact that the gauge parameters are not all independent and a full analysis can be found in \cite{AlvarezV}. It is however irrelevant for the purposes at hand, which are purely tree level.

At the end of the day we are left up with
\bea
&&4 K_{tot}^{U}=k^2 P_2+\left((n-1) m^2-{n-2\over  n^2}k^2\right) P_0^s+\left(m^2-{ n^2-3 n +2\over  n^2}k^2\right) P_0^w+\nonumber\\
&&\left(m^2+{n-2\over  n^2}k^2\right)\sqrt{n-1}P_0^\times+
{k^6\over  M^4}P_1
\eea

So the euclidean propagator is then given in this gauge by

\bea
&&k^2 \Delta= P_2+{M^4\over k^4} P_1-{1\over (n-2) m^2}\bigg\{\left(m^2-{n^2-3n+2\over n^2}k^2\right) P_0^s+\nonumber\\
&&\left((n-1)m^2 -{n-2\over n^2}k^2\right) P_0^w-\left(m^2+{n-2\over n^2}k^2\right)\sqrt{n-1} P_0^\times\bigg\}
\eea

Any coupling of the gravitational fluctuation to an external source $S_{int}=\int d^n x~J_{\m\n} h^{\m\n}$ has to comply with
\be
0=\d S_{int}=\int d^n x~J_{\m\n}\left(\pd_\m \pd_\r\Omega^\r_\m+\pd_\n \pd_\r\Omega^\r_\n+ \omega(x) h_{\m\n}\right)
\ee
which is only possible when the source obeys both $\eta_{\m\n} J^{\m\n}=0$ and $\pd_\m J^{\m\n}=\pd^\n T$. This means that it should be related to some conserved symmetric tensor (which a priori could be different from the usual conserved energy-momentum tensor although we shall prove it to be the same) by 
\be
J_{\m\n}\equiv T_{\m\n}-{1\over n} T \eta_{\m\n}
\ee
 The free energy (or effective action) after the gaussian functional integration just reads
\bea
W[J]&&\equiv {1\over 2}\int d^n x d^n y J^*_{\m\n}(x)~\Delta^{\m\n\r\s}(x,y)~J_{\r\s}(y)=\nonumber\\
&&={1\over 2}(2\pi)^{2n} \int d^n k 
 J_{\m\n}^*(k)~\Delta^{\m\n\r\s}(k)~J_{\r\s}(k)
\eea
and using the easily proved identities
\bea
&&\left(P_1 J\right)_{\m\n}=0\nonumber\\
&&tr\,J P_1 J=0\nonumber\\
&&(P_2)_{\m\n\r\s} J^{\r\s}=T_{\m\n}-{1\over n-1}\theta_{\m\n} T\nonumber\\
&&tr\, \left(J P_2 J\right)= |T_{\m\n}|^2-{1\over n-1}~|T|^2
\eea
it conveys the fact that
\be
W[J]={1\over 2}(2\pi)^{2n} \int d^n k \bigg\{
 J_{\m\n}^*(k)~{1\over k^2}~P_2^{\m\n\r\s}(k)~J_{\r\s}(k)+C(k) |T(k)|^2\bigg\}
 \ee
where
\be
 C(k)=-{1\over (n-1)(n-2)k^2}
\ee
This yields the free energy
\bea\label{free}
&&W[T]={1\over 2}(2\pi)^{2n} \int d^n k 
 {1\over k^2}\left(|J_{\m\n}|^2-{2\over n(n-2)}|T(k)|^2\right)=\nonumber\\
&& ={1\over 2}(2\pi)^{2n} \int d^n k 
 {1\over k^2}\left(|T_{\m\n}|^2-{1\over n-2}|T(k)|^2\right)
 \eea
This result is already implicitly contained in  \cite{AlvarezBGV}. This free energy (when expressed in the second form) is {\em exactly} the same as the prediction of General Relativity, which implies that the low energy physics, and so the low energy empirical tests of gravity, are exactly the same as in General Relativity and convey the same results. It is only in the nonlinear regime that some differences between both theories could possibly be found.

\par

\section{Free energy with external sources in a general background}
In the presence of a vacuum energy, Minkowski space is not an allowed classical solution, so that it is not an appropiate background. Let us  consider instead an arbitrary  background $\bg_{\m\n}$. In this case we will not be able to write down the free energy in a closed form as in the previuos paragraph when we were dealing with flat space. We will rely instead in a detailed analysis of the field equations endowed with arbitary sources. The conclusion will again be that the unimodular EM are equivalent to the GR ones with vanishing cosmological constant. 
\par
Recalling that the unimodular action reads
\be
S_U=\int d^n x \left(g^{1\over n} R+{(n-1)(n-2)\over 4 n^2}{(\nabla g)^2\over g^2}\right)
\ee

The lagrangian, expanded up to second order in the metric perturbation $g_{\mu\nu}=\bg_{\m\n}+\kappa h_{\m\n}$, reads 

\bea
&&L_U\equiv g^{1\over n} R+{(n-1)(n-2)\over 4 n^2}{(\nabla g)^2\over g^2}=\nonumber\\
&&|\bg|^{1\over n}\bigg[\bR+\kappa\left({h\over n}\bR-h^{\a\b} \bR_{\a\b}-\bn^2 h + \bn_\a \bn_\b h^{\a\b}\right)+{\kappa^2\over 4}\bigg\{-2\bn_\b\left(h^{\a\r}\bn^\b h_{\a\r}\right)+\nonumber\\
&&+2\bn_\a\left(-h^{\a\r}\bn_\r h+ 2 h^{\a\r}\bn^\b h_{\r\b}\right)-2 h^{\b\n}\left(-\bn^2 h_{\n\b}+\bn^\r\bn_\n h_{\r\b}+\bn^\r\bn_\b h_{\r\n}-\bn_\b\bn_\n h\right)\nonumber\\
&&-2 \bn_\s h\bn^\s h+2 \bn_\s h\bn_\b h^{\s\b}- 2 \bn_\s h_{\a\n}\bn^\n h^{\a\s}+ \bn_\s h_{\a\n} \bn^\s h^{\a\n}+ 4 h^\b_\a h^{\n\a} \bR_{\n\b}-\nonumber\\
&&{4\over n} h \left(h^{\a\b}\bR_{\a\b}+\bn^2 h-\bn_\a\bn_\b h^{\a\b}\right)+{2\over n}\left({h^2\over n}-h^{\a\b}h_{\a\b}\right)\bR \bigg\}\bigg]+\nonumber\\
&&+{(n-1)(n-2)\over 4 n^2}\bigg\{{(\bn\bg)^2\over \bg^2}+\kappa\left(2\bn_\m h {\bn^\m\bg\over \bg}-h^{\m\n}{\bn_\m\bg\over \bg}{\bn_\n\bg\over\bg}\right)+\nonumber\\
&&\kappa^2\left(-2h^{\m\n}\bn_\m h{\bn_\n\bg\over\bg}+(\bn h)^2-2(h\bn_\m h+h^{\a\b}\bn_\m h_{\a\b}){\bn^\m\bg\over\bg}+h^{\m\a}h^\n_\a {\bn_\m\bg\over\bg}{\bn_\n\bg\over\bg}\right)
\bigg\}
\eea

The only way the linear term can vanish\footnote{Let us dispose of a fine point. Usually the integral of a covariant derivative vanishes because it can be written as
\be
\bn_\m V^\m={1\over \sqrt{|\bg|}}\pd_\m\left(\sqrt{\bg}V^\m\right)
\ee
so that
\be
\int d^n x \sqrt{\bg}\bn_\m V^\m=\int d^n x \pd_\m\left(\sqrt{\bg}V^\m\right)=0
\ee
assuming vanishing physical effects at the boundary. This is not true anymore with the unimodular measure. What can be written instead is

\be
\int d^n x \bg^{1\over n}\bn_\m V^\m={n-2\over n}\int d^n x V^\m \bg^{2-n\over n}~\pd_\m \bg
\ee
} is by restricting either the allowed fluctuations or else the allowed backgrounds through 
\be
{h\over n}\bR-h^{\a\b} \bR_{\a\b}-\bn^2 h + \bn_\a \bn_\b h^{\a\b}+2\bn_\m h {\bn^\m\bg\over \bg}-h^{\m\n}{\bn_\m\bg\over \bg}{\bn_\n\bg\over\bg}=0
\ee
Which for maximally symmetric backgrounds, in which $\bR_{\m\n}=-{2\l\over n-2}\bg_{\m\n}$, reads
\be
\bn^2 h - \bn_\a \bn_\b h^{\a\b}+2\bn_\m h {\bn^\m\bg\over \bg}-h^{\m\n}{\bn_\m\bg\over \bg}{\bn_\n\bg\over\bg}=0
\ee
A simple solution consists in restricting the background  to be unimodular by itself, id est,
\be
\bg=1
\ee
in which case the offending terms again either vanish or else behave as total derivatives.

To summarize, once a unimodular background is chosen the linear term is just the equation of motion for the background field
\begin{align}
h^{\mu\nu}\left( \bar{R}_{\mu\nu}-\frac{1}{n}\bar{R}\bar{g}_{\mu\nu}\right)=0
\end{align}

which is the vacuum field equation of the total non-linear unimodular theory, in which Bianchi identities force the scalar curvature to be constant.
\par
We are finally ready for our analysis of the EM of both theories around arbitrary backgrounds $\bar{g}_{\m\n}$ and $\hat{g}_{\m\n}$.
Remember that  the lagrangians for the U theory and for GR with cosmological constant\footnote{
In the full nonlinear theory the cosmological constant is included in an arbitrary energy momentum tensor. In the linear approximation this is not the case.
}, both expanded up to second order in linear perturbations, are

\begin{align}
L_{U}&=\frac{n+2}{4n^{2}}\bar{\nabla}^{\mu}h\bar{\nabla}_{\mu}h-\frac{1}{n}\bar{\nabla}_{\mu}h\bar{\nabla}^{\rho}h^{\mu}_{\rho}+\frac{1}{2}\bar{\nabla}_{\mu}h^{\mu\rho}\bar{\nabla}_{\nu}h^{\nu}_{\rho}-\\
\nonumber &-\frac{1}{4}\bar{\nabla}_{\mu}h^{\nu\rho}\bar{\nabla}^{\mu}h_{\nu\rho}-\bar{R}_{\nu\beta}h^{\beta}_{\alpha}h^{\nu\alpha}+\frac{1}{n}h\bar{R}_{\alpha\beta}h^{\alpha\beta}-\frac{\bar{R}}{2}\left( \frac{h^{2}}{n^{2}}-\frac{1}{n}h^{\alpha\beta}h_{\alpha\beta}\right)\\
\nonumber \\
L_{GR\lambda}&=\frac{1}{4}\hat{\nabla}^{\mu}h\hat{\nabla}_{\mu}h-\frac{1}{2}\hat{\nabla}_{\mu}h\hat{\nabla}^{\rho}h^{\mu}_{\rho}+\frac{1}{2}\hat{\nabla}_{\mu}h^{\mu\rho}\hat{\nabla}_{\nu}h^{\nu}_{\rho}-\\
\nonumber &-\frac{1}{4}\hat{\nabla}_{\mu}h^{\nu\rho}\hat{\nabla}^{\mu}h_{\nu\rho}-\hat{R}_{\nu\beta}h^{\beta}_{\alpha}h^{\nu\alpha}+\frac{1}{2}h\hat{R}_{\alpha\beta}h^{\alpha\beta}-\frac{\hat{R}+2\lambda}{2}\left( \frac{h^{2}}{4}-\frac{1}{2}h^{\alpha\beta}h_{\alpha\beta}\right)
\end{align}

\vspace{0.3cm}
Assuming both backgrounds to be of maximally symmetric spaces $R_{\mu\nu}=-\frac{2\lambda}{n-2}g_{\mu\nu}$, they reduce to 
\begin{align}
L_{U}&=\frac{n+2}{4n^{2}}\bar{\nabla}^{\mu}h\bar{\nabla}_{\mu}h-\frac{1}{n}\bar{\nabla}_{\mu}h\bar{\nabla}^{\rho}h^{\mu}_{\rho}+\frac{1}{2}\bar{\nabla}_{\mu}h^{\mu\rho}\bar{\nabla}_{\nu}h^{\nu}_{\rho}-\frac{1}{4}\bar{\nabla}_{\mu}h^{\nu\rho}\bar{\nabla}^{\mu}h_{\nu\rho}\\
\nonumber \\
L_{GR\lambda}&=\frac{1}{4}\hat{\nabla}^{\mu}h\hat{\nabla}_{\mu}h-\frac{1}{2}\hat{\nabla}_{\mu}h\hat{\nabla}^{\rho}h^{\mu}_{\rho}+\frac{1}{2}\hat{\nabla}_{\mu}h^{\mu\rho}\hat{\nabla}_{\nu}h^{\nu}_{\rho} -\frac{1}{4}\hat{\nabla}_{\mu}h^{\nu\rho}\hat{\nabla}^{\mu}h_{\nu\rho}-\nonumber\\
&-\frac{\lambda}{2}\left( \frac{h^{2}}{2}-h_{\alpha\beta}h^{\alpha\beta}\right)
\end{align}

Sources for both theories can be introduced in the usual way by a linear coupling. In the case of GR, the source is just the usual symmetric energy-momentum tensor $T_{\mu\nu}$ while for the unimodular theory is its traceless source  $J_{\mu\nu}$ which obeys $\nabla_\m J^{\m\n}=\nabla^\n T$. 

\vspace{0.4cm}
The equations of motion of the unimodular theory, dubbed EMU, then read

\bea
&&EMU\equiv \frac{n+2}{2n^{2}}\bar{g}_{\mu\nu}\bar{\nabla}^{2}h-\frac{1}{2}\bar{\nabla}^{2}h_{\mu\nu}-\frac{1}{n}\bar{\nabla}_{\alpha}\bar{\nabla}_{\beta}h^{\alpha\beta}\bar{g}_{\mu\nu}-\nonumber\\
&& \frac{1}{n}\bar{\nabla}_{\mu}\bar{\nabla}_{\nu}h + \frac{1}{2}\bar{\nabla}_{\mu}\bar{\nabla}_{\alpha}h^{\alpha}_{\nu}+\frac{1}{2}\bar{\nabla}_{\nu}\bar{\nabla}_{\alpha}h^{\alpha}_{\mu}=J_{\mu\nu}
\eea
whereas the equations of motion of general relativity, EMGR, are
\bea
&&EMGR\equiv \frac{1}{2}\hat{\nabla}^{2}h\hat{g}_{\mu\nu}-\frac{1}{2}\hat{\nabla}^{2}h_{\mu\nu}-\frac{1}{2}\hat{\nabla}_{\alpha}\hat{\nabla}_{\beta}h^{\alpha\beta}\hat{g}_{\mu\nu}-\frac{1}{2}\hat{\nabla}_{\mu}\hat{\nabla}_{\nu}h+\nonumber\\
&& \frac{1}{2}\hat{\nabla}_{\mu}\hat{\nabla}_{\alpha}h^{\alpha}_{\nu}+\frac{1}{2}\hat{\nabla}_{\nu}\hat{\nabla}_{\alpha}h^{\alpha}_{\mu}= \lambda \left(\frac{h}{2}\hat{g}_{\mu\nu}-h_{\mu\nu}\right)+T_{\mu\nu}
\eea

At this point we should remember the result advertised in the introduction on the equivalence of the unimodular theory with GR with an undetermined cosmological constant. We have already analyzed in the previous section fluctuations around a flat background and found full equivalence with GR with vanishing cosmological constant. In order to make sure that this result is not an artifact of the flat background, it is worth to repeat the analysis in this more general setting.
\par
To make things easy, we can derive two first integrals from the equations of motion. The first one,  $I_{GR}$, by taking the trace of the EMGR:

\be
I_{GR}\equiv \hat{\nabla}^{2}h-\hat{\nabla}_{\alpha}\hat{\nabla}_{\beta}h^{\alpha\beta}-\lambda h=\frac{2}{n-2}T\\
\ee
whereas the second one stems from taking the covariant divergence of the EMU
\bea
&&I_U\equiv \frac{2-n}{2n}\bar{\nabla}_{\nu}\left( \bar{\nabla}_{\alpha}\bar{\nabla}_{\beta}h^{\alpha\beta}-\frac{1}{n}\bar{\nabla}^{2}h+\frac{1}{n}T\right)=0 \nonumber\\
&&\Rightarrow   \bar{\nabla}_{\alpha}\bar{\nabla}_{\beta}h^{\alpha\beta}-\frac{1}{n}\bar{\nabla}^{2}h+\frac{1}{n}T=\Gamma
\eea

where $\Gamma$ is an arbitrary constant.
Now let us assume that the background metric is the same for both theories. Since we are trying to check if they are equivalent, this is a reasonable ansatz, so we set $\bar{g}_{\mu\nu}=\hat{g}_{\mu\nu}$. After that, we look for a field redefinition of the form $h_{\mu\nu}=H_{\mu\nu}+a H\bar{g}_{\mu\nu}$  that would take one theory into the other, with the possible addition of terms proportional to the first integrals which are zero by the use of the equations of the motion. 
This is equivalent to a search for  constants $a$, $C_{1}$ $C_{2}$ and $\Gamma$ such that
\be
EMGR\left(H_{\m\n}+ a H \bg_{\m\n}\right)+ C_2 I_{GR}\left(H_{\m\n}+ a H \bg_{\m\n}\right)= EMU\left(H_{\m\n}\right)+C_1 I_U\left(H_{\m\n}\right)
\ee

\bea
&&\bar{\nabla}^{2}H \bar{g}_{\mu\nu}\left( \frac{n+2}{2n^{2}}-\frac{C_{1}}{n}-\frac{1}{2}-a\left( \frac{n}{2}-1\right)-C_{2}(1+na-a) \right)+\nonumber\\
&&+\bar{\nabla}_{\alpha}\bar{\nabla}_{\beta}H^{\alpha\beta}\bar{g}_{\mu\nu}\left( C_{1}+C_{2}+\frac{1}{2}-\frac{1}{n}\right)+\bar{\nabla}_{\mu}\bar{\nabla}_{\nu}H\left( \frac{1}{2}-\frac{1}{n}-a\left( 1-\frac{n}{2}\right)\right)+\nonumber\\
&&+H\bar{g}_{\mu\nu}\lambda\left(\frac{1}{2}+a\left( \frac{n}{2}-1\right)+C_{2}(1+na)\right)+\bar{g}_{\mu\nu}\left(	T\left(  \frac{2C_{2}}{n-2}+\frac{C_{1}+1}{n}\right)-C_1 \Gamma\right)+\nonumber\\
&&+T_{\m\n}-J_{\m\n}-
\lambda H_{\mu\nu}=0
\eea

The system of equations obtained by demanding  every factor to be zero is only compatible if the cosmological constant $\lambda$ vanishes. In that case, the solution of the system is simply

\bea
&&a=-\frac{1}{n}\nonumber\\
&& C_{1}+C_{2} = \frac{2-n}{2n}\nonumber\\
&& \Gamma = \left( \frac{n^{2}(2C_{2}-1)+n(4C_{2}+4)-4}{2n^{2}(n-2)}\right) T
\eea
To summarize,
\be
EMGR\left(H_{\m\n}-{1\over n} H \bg_{\m\n}\right) -{n-2\over 2n}I_{GR}\left(H_{\m\n}-{1\over n} H \bg_{\m\n}\right)=EMU\left(H_{\m\n}\right)\left.\right|_{J_{\m\n}=T_{\m\n}-{1\over n} T \bg_{\m\n}}
\ee
The physical meaning of what we have proved is that the unimodular EMU are a consequence of EMGR  when $\l=0$ {\em only}; actually it is the subsector corresponding to
\be
h^{GR}_{\m\n}=h^U_{\m\n}-{1\over n} h^U g_{\m\n}
\ee
We insist that this is {\em not} a field redefinition; is a truncation of GR such that $h^{GR}=0$. There is no way to build the inverse map from EMU to EMGR. Given the fact that
\be
h=0
\ee
is a (partial) algebraic gauge fixing (which does not need ghosts), this shows that, at the level of the equations of motion, the unimodular theory is a truncation of GR with vanishing cosmological constant, and with the source reduced to the traceless part of the GR source. It is perhaps worth remarking  that this does not follow necessarily from the fact that the lagrangian is so obtained (gauge conditions can only be used {\em after} the EM are derived).

\section{Conclusions}

The main conclusion of our work is that, although the only allowed source of the gravitational field in the unimodular theory is just the traceless piece of the energy-momentum tensor
\be
J_{\m\n}\equiv T_{\m\n}-{1\over n} T g_{\m\n}
\ee
the EM of the arbitrary quadratic fluctuations in the unimodular theory are equivalent to the corresponding EM of  General Relativity with the full source $T_{\m\n}$ and this holds before any gauge fixing. 

\par It could be naively thought that this result is just a trivial consequence of the fact that
\be
g^E_{\m\n}\equiv g^{-{1\over n}}g_{\m\n}
\ee
Actually this a delusion. The EM corresponding to the lagrangian in which a point transformation has been preformed are equivalent to the initial ones only \cite{Arnold} if the transformation is invertible, which means that the  jacobian must be nonvanishing, which is not the case. This means that the correct way of looking at the unimodular theory is as a presumably consistent truncation of general relativity, inequivalent to it, and one that implies that the cosmological constant must vanish. In some sense this is not so different as the way superstrings can be understood as a GSO projection of the NSR string.
\par 

In the particular case that we are interested in fluctuations with respect to a flat background we were able to prove a stronger result  namely, that
the {\em full free energy} produced by {\em arbitrary} sources (not only static ones) is {\em identical} to the one predicted by general relativity. 
\par
 An important question  is  whether this truncation will survive quantum corrections. We hope to report of this in the near future.

\newpage
\appendix
\section{Barnes-Rivers projectors in momentum space.}

Let us briefly state our notation (the same as in \cite{AlvarezBGV}). We start with the longitudinal and transverse projectors
\bea
&&\theta_{\a\b}\equiv\eta_{\a\b}-{k_\a k_\b\over k^2}\nonumber\\ 
&&\omega_{\a\b}\equiv {k_\a k_\b\over k^2}
\eea
They obey
\bea
&&\theta+\omega\equiv \theta_\m^\n+\omega_\m^\n=\d_\m^\n\equiv 1\nonumber\\
&&\theta^2\equiv \theta_\a^\b\theta_\b^\g=\theta_\a^\g\equiv \theta\nonumber\\
&&\omega^2\equiv \omega_\a^\b \omega_\b^\g=\omega_\a^\g\equiv \omega
\eea
as well as
\bea
&& tr~\theta=n-1\nonumber\\
&&tr~\omega=1
\eea
The four-indices projectors are
\bea
&&P_2\equiv {1\over 2}\left(\theta_{\m\r}\theta_{\n\s}+\theta_{\m\s}\theta_{\n\r}\right)-{1\over n-1}\theta_{\m\n} \theta_{\r\s}\nonumber\\
&&P_1\equiv{1\over 2}\left(\theta_{\m\r}\omega_{\n\s}+\theta_{\m\s}\omega_{\n\r}+\theta_{\n\r}\omega_{\m\s}+\theta_{\n\s}\omega_{\m\r}\right)\nonumber\\
&&P_0^s\equiv {1\over n-1}\theta_{\m\n}\theta_{\r\s}\nonumber\\
&&P_0^w\equiv \omega_{\m\n}\omega_{\r\s}\nonumber\\
&&P_0^{sw}\equiv{1\over \sqrt{n-1}}\theta_{\m\n}\omega_{\r\s}\nonumber\\
&&P_0^{ws}\equiv{1\over \sqrt{n-1}}\omega_{\m\n}\theta_{\r\s}
\eea
They obey
\bea
&&P_i^a P_j^b=\d_{ij}\d^{ab} P_i^b\nonumber\\
&&P_i^a P_j^{bc}=\d_{ij}\d^{ab}P_j^{ac}\nonumber\\
&&P_i^{ab} P_j^c=\d_{ij}\d^{bc} P_j^{ac}\nonumber\\
&&P_i^{ab} P_j^{cd}=\d_{ij}\d^{bc}\d^{ad} P_j^a
\eea
as well as
\bea
&&tr~P_2\equiv \eta^{\m\n} (P_2)_{\m\n\r\s}=0\nonumber\\
&&tr~P_0^s=\theta_{\r\s}\nonumber\\
&&tr~P_0^w=\omega_{\r\s}\nonumber\\
&&tr~P_1=0\nonumber\\
&&tr~P_0^{sw}=\sqrt{n-1}~\omega_{\r\s}\nonumber\\
&&tr~P_0^{ws}={1\over \sqrt{n-1}}~\theta_{\r\s}\nonumber\\
&&P_2+P_1+P_0^w+P_0^s={1\over 2}\left(\d_\m^\n \d_\r^\s+\d_\m^\s \d_\r^\n\right)
\eea

Any symmetric operator can be written as
\be
K= a_2 P_2 + a_1 P_1 + a_w P_0^w + a_s P_0^s + a_\times P_0^\times
\ee
(where $P_0^\times\equiv P_0^{ws}+P_0^{sw}$).
Then
\be
K^{-1}={1\over a_2}P_2+{1\over a_1} P_1 +{a_s\over a_s a_w - a_\times^2}P_0^w+{a_w\over a_s a_w - a_\times^2}P_0^s-{a_\times\over a_s a_w - a_\times^2}P_0^\times
\ee

Sometimes the action of those projectors on tracefree tensors  is  needed. Defining the trecefree projector
\be
\left(P_{tr}\right)_{\r\s}\,^{\l\d}\equiv {1\over 2}\left(\d_\r^\l \d_\s^\d+\d_\r^\d \d_\s^\l\right)-{1\over n}\eta_{\r\s}\eta^{\l\d}
\ee

It is a fact that
\bea
&&\left(P_2\right)_{\m\n}\,^{\r\s}\left(P_{tr}\right)_{\r\s}\,^{\l\d}=P_2\nonumber\\
&&P_0^s P_{tr}=P_0^s-{n-1\over n}P_0^s-{\sqrt{n-1}\over n}P_0^{sw}\nonumber\\
&&P_0^w P_{tr}=P_0^w-{\sqrt{n-1}\over n}P_0^{ws}-{1\over n}P_0^w\nonumber\\
&&P_1 P_{tr}=P_1\nonumber\\
&&P_0^{sw} P_{tr}=P_0^{sw}-{\sqrt{n-1}\over n}P_0^{ws}-{1\over n}P_0^w\nonumber\\
&&P_0^{ws} P_{tr}=P_0^{ws}-{\sqrt{n-1}\over n} P_0^{sw}-{n-1\over n} P_0^s
\eea

\newpage
\section*{Acknowledgments}
One of us (EA) is grateful for  e-mail exchange with Stanley Deser. He is also indebted to the former transverse collaborators Diego Blas, Jaume Garriga and Enric Verdaguer. This work has been partially supported by the European Union FP7  ITN INVISIBLES (Marie Curie Actions, PITN- GA-2011- 289442)and (HPRN-CT-200-00148) as well as by FPA2009-09017 (DGI del MCyT, Spain) and S2009ESP-1473 (CA Madrid).


\end{document}